\documentstyle[aps,prl,amsfonts,epsfig,preprint]{revtex}

\begin{document}

\title{Fluxes of cosmic rays: A delicately balanced stationary state}

\author{Constantino Tsallis,$^{1}$%
\thanks{tsallis@cbpf.br}
Joao C. Anjos,$^{1}$ 
Ernesto P. Borges$^{1,2}$ \\
$^1$Centro Brasileiro de Pesquisas Fisicas, 
Rua Xavier Sigaud 150 \\
22290-180 Rio de Janeiro-RJ, Brazil\\
$^2$Escola Politecnica,
Universidade Federal da Bahia,
Rua Aristides Novis 2 \\
40210-630 Salvador-BA, Brazil}

\maketitle

\begin{abstract}
The analysis of cosmic rays fluxes as a function of energy reveals a {\it knee} slightly below
$10^{16}$ eV and an {\it ankle} close to $10^{19}$ eV. 
Their physical origins remain up to now quite enigmatic; in particular,  
no elementary process is known which occurs at energies close to $10^{16}$ eV.
We propose a phenomenological approach along the lines of nonextensive
statistical mechanics, a formalism which contains Boltzmann-Gibbs statistical
mechanics as a particular case. 
The knee then appears as a crossover between two fractal-like thermal
regimes, the crossover being caused by process occurring at energies 
ten million times lower than that of the knee, in the region of the 
quark hadron transition ($\simeq 10^{9}$ eV). 
This opens the door to an unexpected standpoint for further clarifying 
the phenomenon.

$ $


\end{abstract}

\bigskip

Cosmic rays fascinate since long. They provide galactic, extragalactic and
cosmological information, related to recent or very old events concerning
various sources, going back to the early times of the universe \cite{cronin}. 
They reflect all types of elementary process and interactions, and are
associated with phenomena of very different space and time scales. 
The complete physical scenario still remains quite enigmatic, although a 
variety of specific mechanisms for acceleration and propagation have been 
advanced along the years for various energy regions. The most known of these 
mechanisms is the Fermi one \cite{fermi}, which addresses acceleration in 
magnetized turbulent plasma, but many others have been advanced 
\cite{peters,lagage,drury,wdowczyk,ptuskin,dova,horandel,axford} in relation 
with the {\it knee} and energies below it; for the energies beyond these, 
for the {\it ankle}, as well as for general reviews, see 
\cite{nagano,sigl,malkov}.

\begin{figure}[htb!]
 \begin{center}
 \epsfig{figure=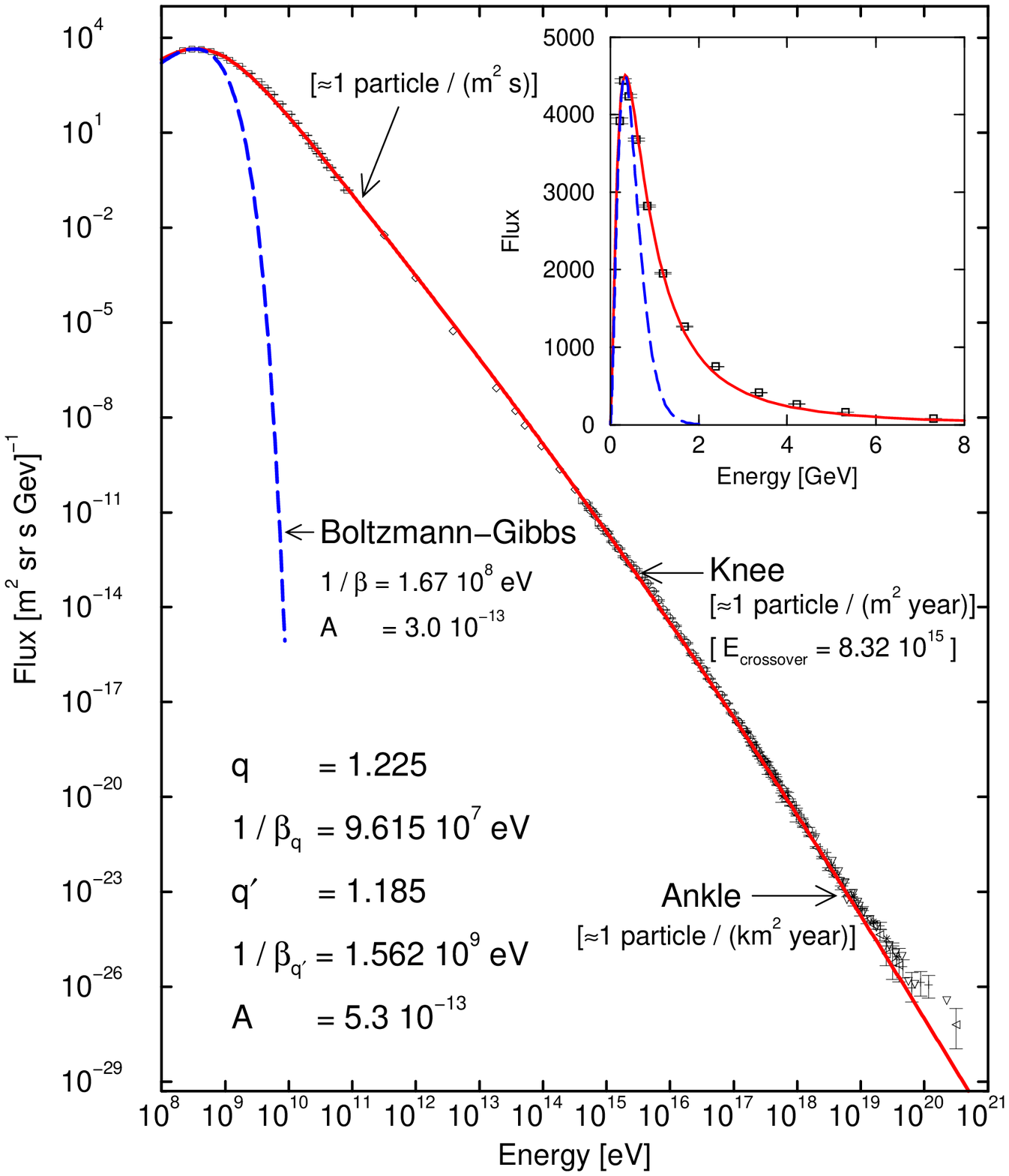,width=0.7\textwidth,clip=}
 \end{center}
 \caption[]
 {Energy dependence of the fluxes of cosmic rays. 
 Experimental error bars are indicated whenever available. 
 The continuous curve is the one we obtain within the present phenomenological 
 approach. The dashed curve is an optimized BG one (even at relatively low 
 energies it fails by very many decades). 
 The knee corresponds to $E_{crossover}$. 
 Inset: Linear-linear representation of the low energy fluxes.}
 \label{fig:fluxes}
\end{figure}

Through various types of detectors, the flux of cosmic rays at the top of 
the Earth atmosphere has been measured \cite{swordy}
and varies from $10^4$ down to $10^{-29} [m^2\;sr\; s\; \mbox{GeV}]^{-1}$  
for energies increasing from $10^8$  up to near $10^{21}$ eV: See Fig. 
\ref{fig:fluxes}. 
This distribution (which spans 13 decades in energy and 33 decades in flux!) 
is {\it not} exponential, hence it does not correspond to Boltzmann-Gibbs (BG) 
statistical mechanics thermal equilibrium. Consistently, even at a 
phenomenological level, i.e., without specifying any concrete model or 
mechanism, this problem represents a challenge. 
This is the one we address here. We shall use a point of view based on a 
current generalization of Boltzmann-Gibbs statistical mechanics, referred to 
as nonextensive statistical mechanics, we shall briefly describe later on. 
The first step will be to remark that the fluxes of cosmic rays in general, 
and the studies of the``knee" and the ``ankle" in particular, involve phenomena
such as turbulence (see, for instance, \cite{candia}), anomalous diffusion and 
fractality (see, for instance, \cite{lagutin}), self-organized criticality 
(see, for instance, \cite{SOC}), long-range interactions (classical and quantum 
gravitation), among other complex phenomena (such as, for example, possible 
nonmarkovianity \cite{bediaga}). It is precisely such phenomena that constitute
the scope of nonextensive statistical mechanics; for turbulence and related 
matters see, for instance, \cite{turbulence2,turbulence3,turbulence6}; 
for anomalous diffusion and fractality see, for instance, 
\cite{zanette,tsallislevy,bukman,wilkrapidity,baldovinrobledo,borgesetal,%
weinstein};  
for self-organized criticality see, for instance, \cite{tamarit}; 
for long-range interactions see, for instance, 
\cite{anteneodo,latorarapisardatsallis,cabral}. 

Let us now first briefly review the usual, BG thermostatistics.
If we optimize under appropriate constraints the BG 
$S=-k \sum_i p_i \ln p_i$ ($k\equiv $ Boltzmann constant;
$\{p_i\} \equiv$ microscopic probabilities) we obtain the celebrated equilibrium
distribution $p_i =\frac{e^{-\beta E_i}}{Z} \propto e^{-\beta E_i}$ ($\beta \equiv 1/kT$, $E_i\equiv$ energy of the {\it i-th} state; 
$Z \equiv \sum_j e^{-\beta E_j}\equiv$ partition function). 
Excepting for the trivial normalizing factor $1/Z$, 
this distribution can alternatively be obtained as the solution of 
the  linear differential equation 
\begin{equation}
dp_i/dE_i = -\beta p_i \;.
\end{equation}
 
In order to deal with a variety of thermodynamically anomalous systems, 
a more general formalism, nonextensive statistical mechanics, was introduced in 1988 \cite{tsallis1,curadotsallis,tsallis3,kaniadakis}.
It is based on the generalized entropic form
$S_q=k(1-\sum_i p_i^q)/(q-1)$ ($q \in \mathbb{R}$ and $S_1=S$). 
Its optimization under appropriate constraints yields \cite{tsallis2} 
a power-law, $p_i \propto [1-(1-q)\beta_q E_i]^{\frac{1}{1-q}} \equiv e_q^{-\beta_q E_i}$ (definition), which 
recovers the BG weight for $q=1$ ($\beta_1\equiv \beta$). 
As usual, $kT_q \equiv 1/\beta_q$ characterizes the conveniently averaged 
energy. This anomalous equilibrium-like distribution can be alternatively 
obtained (excepting for the normalizing factor) by 
solving the nonlinear differential equation 
\begin{equation}
dp_i/dE_i = -\beta_q p_i^q \;.
\end{equation} 
This generalized weight naturally emerges in ubiquitous problems such as 
fully developed turbulence \cite{turbulence2,turbulence3,turbulence6} 
(which is  relevant for  the Fermi mechanism, and most probably for others as 
well), electron-positron annihilation \cite{bediaga}, 
motion of {\it Hydra viridissima} \cite{arpita}, 
long-range many-body Hamiltonians \cite{latorarapisardatsallis}, 
among many others.

We now use the above differential equation path in order to further
generalize the anomalous equilibrium distribution, in such a way as to 
have a crossover from anomalous ($q \ne 1$) to normal ($q=1$)
thermostatistics, while increasing the energy. We consider then the 
differential equation 
\begin{equation}
dp_i/dE_i = -\beta_1 p_i -(\beta_q-\beta_1) p_i^q \;, 
\end{equation}
whose solution is 
$p_i \propto  [1-\frac{\beta_q}{\beta_1}+
\frac{\beta_q}{\beta_1}\;e^{(q-1)\beta_1 E_i}]^{-\frac{1}{q-1}}$. 
The crossover typically occurs for $q>1$ and $\beta_1 \ll \beta_q$, the
distribution being anomalous at low energies and BG at high
energies. It is undoubtedly interesting to notice that this differential 
equation precisely coincides, for $q=2$, with the 
heuristic one that in 1900 led Planck to the discovery of the 
black-body radiation law and ultimately to quantum mechanics \cite{planck}.

Finally, by doing one more step along the same direction, we can further
generalize the differential equation, now becoming 
\begin{equation}
dp_i/dE_i = -\beta_{q^{\prime}} p_i^{q^{\prime}} 
-(\beta_q-\beta_{q^{\prime}}) p_i^q \;.
\end{equation} 
This manner of writing the coefficient of the $p_i^q$ term has the advantage
of recovering the simplest generalization of the BG distribution by
considering $\beta_{q^\prime}=\beta_q$ or  $\beta_{q^\prime} = 0$ or even
$q^\prime=q$. For $1<q^{\prime}<q$ and
$\beta_{q^{\prime}} \ll \beta_q$, a crossover occurs at
\begin{equation}
E_{crossover}=[(q-1)\beta_q]^{\frac{q^\prime-1}{q-q^\prime}}    /  
[(q^\prime-1)\beta_{q^\prime}]^{\frac{q-1}{q-q^\prime}} \;.
\end{equation}
For $E \ll E_{crossover}$ we have an anomalous 
distribution characterized by $(q,\beta_q)$ (namely 
$p_i \propto e_q^{-\beta_q E_i}$), whereas for $E \gg E_{crossover}$ 
we have a different anomalous distribution characterized by 
$(q^{\prime},\beta_{q^{\prime}})$ 
(namely $p_i \propto e_{q^\prime}^{-\beta_{q^\prime} E_i}$). 
The exact solution of the above differential equation (the most general 
one considered here) is given by $p_i \propto f(E_i)$ where $f^{-1}(x)$ is an
explicit monotonic function of $x$ involving hypergeometric functions 
(see Ref. \cite{bemski} 
for details). Interestingly enough, this precise 
solution arrives in the discussion of the re-association of $CO$ molecules
in Myoglobin \cite{bemski}, 
where time plays a role very analogous to the one
played by energy in our cosmic rays problem. This time-energy analogy is 
not surprising after all if we take into account that, in the history of 
the universe after the big-bang, the time scale reflects  
the energy scale, as discussed in detail in Ref. \cite{kolbturner}.

The flux $\Phi(E)$ can be obtained straightforwardly from $p_i \propto f(E_i)$
by calculating the density of states $\omega(E)$. In the ultrarelativistic 
limit $E \propto |{\bf p}|$ (${\bf p}$ being the momentum), which we adopt 
here for simplicity given the high values of the involved energies, the 
density of states of an ideal gas in three dimensions is given by  
$\omega(E) \propto E^2$, hence $\Phi(E)=AE^2f(E)$, where $A$ is a normalizing
factor (and where red shift effects have been neglected). With this expression 
we fit the observational data and obtain the results displayed in 
Fig.~\ref{fig:fluxes}. 
As we can see, the agreement is quite remarkable.

Our summarizing comments are:

(i) The high quality agreement over so many decades, including crossovers 
between different regimes, suggests that the phenomenological approach is 
correct, and specific models clarifying the various physical mechanisms that 
are involved should essentially satisfy it;

(ii) The deep explanation of the knee might well be found at energies 
extremely lower (ten million times lower, in fact), basically at energies
related to the characteristic temperatures obtained from the fitting, namely 
$9.615\;10^7$ eV (energy comparable to the pion mass) and  
$1.562 \;10^9$ eV (energy corresponding to the quark-hadron transition 
\cite{kolbturner}, as well as to the proton mass), and given the fact that 
$q$ and $q^{\prime}$ differ by only $3 \%$ (which in log-log representation 
asymptotically determines two almost parallel straight lines which very slowly 
approach to each other and intersect at $E=E_{crossover}$). 
The existence of two thermostatistical regimes, respectively related to
($q,\beta_q$) and ($q^\prime,\beta_{q^\prime}$), could correspond to two 
different mechanisms of acceleration/propagation, for instance related to 
galactic and extra-galactic contributions. It is worthy stressing at this point
that the location of the knee emerges here through Eq. (5), i.e., using only 
the basic four phenomenological parameters 
$(q,\beta_q, q^\prime,\beta_{q^\prime})$, and not by introducing an extra 
parameter on top of the previous ones (whose role would be to fix the knee at 
the right value);

(iii) Since the entropic index $q$ is known to reflect (multi) fractality 
\cite{lyra,moura}, the present results strongly suggest that either the 
generation or the transport (or both) of cosmic rays occur in scale 
invariant media, which is consistent with Ref. \cite{lagutin};

(iv) With the better observational statistics (i.e., with refined precision) 
at these very high energies, expected from the Pierre Auger Observatory 
(or from similar projects), it might happen that the ankle disappears; if it 
does not, then it is probably associated with strongly nonstationary phenomena 
(perhaps related to memory effects from early cosmic stages), 
certainly out from the present thermodynamical description --- 
such study would eventually clarify the physical meaning, if any, of the GZK 
feature \cite{greisen,ZK,olinto1,olinto2} (see also \cite{fengshapere}).

(v) The Kaskade collaboration \cite{horandel} has shown the relevance of the 
various constituents which compose cosmic rays. In order to take this into 
account, the present approach could be further improved by using the 
grand-canonical instead of the canonical ensemble used here.

(vi) The average energy 
$\langle E\rangle\equiv\int_0^\infty dE\,E\,\Phi(E)/\int_0^\infty dE\,\Phi(E)$ 
has been calculated, within the present approach, to be 
$\langle E \rangle \simeq 2.489\,\mbox{GeV}$. Any connection of this value with
other cosmological or astrophysical quantities is of course very welcome.\\

Stimulating remarks from I. Bediaga, L. Galgani, R.C. Shellard and L. Masperi 
are acknowledged. 
Partial support from CNPq, PRONEX, CAPES and FAPERJ (Brazilian agencies) 
is also acknowledged.

\end{document}